\documentstyle[prl,aps,multicol]{revtex}
\input{epsf}
\begin{document}
\draft
\title{Magnetic Field Effect for Two Electrons 
in a Two Dimensional Random Potential}  

\author{Giuliano Benenti$^{(a,b,c)}$ and Dima L. Shepelyansky$^{(a)}$}

\address{$^{(a)}$Laboratoire de Physique Quantique, UMR 5626 du CNRS, 
Universit\'e Paul Sabatier, 31062 Toulouse Cedex 4, France}
\address{$^{(b)}$International Center for the Study of Dynamical 
Systems, \\ 
Universit\`a degli Studi dell'Insubria, via Valleggio 11, 22100 Como, Italy}  
\address{$^{(c)}$Istituto Nazionale di Fisica della Materia, 
Unit\`a di Milano, via Celoria 16, 20133 Milano, Italy} 

\date{November 28, 2000}

\maketitle

\begin{abstract}
We study the problem of two particles with Coulomb repulsion 
in a two-dimensional disordered 
potential in the presence of a magnetic field. 
For the regime, when without interaction  all states are well localized,
it is shown that above a critical excitation energy 
electron pairs become delocalized by interaction. The transition
between the localized and delocalized regimes  
goes in the same way as  the metal-insulator transition
at the mobility edge  in  the three dimensional  
Anderson model with broken time reversal symmetry.  
\end{abstract}

\begin{multicols}{2}

\narrowtext

\section{Introduction}
\label{intro}

The interplay of disorder and interactions in electronic systems 
is a central problem in condensed matter physics \cite{loc99}. 
Two-dimensional (2D) systems are of a particular interest, since 
the scaling theory of localization \cite{abrahams} predicts that 
noninteracting electrons are always localized in a 2D disordered 
potential, while a metal-insulator transition has been reported in 
transport measurements with 2D electron and hole gases 
\cite{kravchenko}. 
The study of this many-body problem is very complicated, 
both for analytical and numerical analysis. 
It is therefore highly desirable to have some relatively simple 
models which could be solved and would lead to a better understanding 
of the effects of interactions in the presence of disorder.    
The problem of two interacting particles (TIP) in a random potential 
has received much attention in the last few years. It has been 
shown \cite{dima} that the TIP can propagate coherently on a 
length $l_c$ which is much larger than the one-particle localization length 
$l_1$ that  can lead to an enhancement of transport \cite{imry}. 
This problem has been studied recently by different groups in one  
\cite{dima,imry,JLP,oppen,song,schreiber,moriond} and 
two dimension \cite{cuevas,pollak,dima2d} and it has been understood 
that the pair delocalization is related to the enhancement of 
interaction in systems with complex, chaotic  eigenstates. The delocalization  
factor is determined by the density of two-particle states 
coupled by the interaction, $\rho_2$, and by the interaction induced 
transition rate, $\Gamma_e$. At $\kappa_e= \Gamma_e \rho_2\sim 1$ the 
interaction matrix elements become comparable with the two-particle 
level spacing and the collisions between particles 
give a strong increase of the ratio $l_c/l_1$.

Most studies of the TIP problem have considered a short range interaction. 
In this case, when two particles are localized at a distance $R\gg l_1$,
the overlap of their wavefunctions is exponentially small, and such 
states are localized much in the same way as in the noninteracting case. 
On the other hand, when the 
average distance between particles is larger than $l_1$, the screening 
of charges in the problem with
a given charge density is problematic. Due to that
for such a regime it is natural to consider the bare Coulomb 
interaction in the simple TIP problem.   
Recently, it has been shown \cite{dima2d} that the Coulomb repulsion can 
delocalize two particles (electrons) in a two-dimensional disordered lattice, 
even if the particles are separated by a distance $R\gg l_1$. 
The delocalization of two-electron states takes place in a way 
similar to the single particle Anderson transition in three dimensions. 
Indeed, the pair center of mass moves in the 2D plane, while electrons rotate 
around it that gives an effective third dimension. The rotation goes on 
a ring of width $l_1$ and radius $R\propto l_1^{4/3}$ fixed by energy 
conservation. As  a result the two-particle states are delocalized 
for excitation energies $\epsilon>\epsilon_c\propto l_1^{-4/3}$ 
($\kappa_e>1$) \cite{dima2d}. 
This expectations have been confirmed numerically \cite{dima2d}   
by a study of the level spacing statistics $P(s)$, which displays 
a transition from the
Poisson distribution (for $\epsilon<\epsilon_c$) to 
the Wigner-Dyson distribution (for $\epsilon>\epsilon_c$).
It was also found that at  a critical point ($\epsilon=\epsilon_c$) 
$P(s)$ statistics  is close to the distribution found in the 3D Anderson 
model at the mobility edge \cite{zhar2,braun}.  

In this paper we consider, for the first time, the effect of a magnetic 
field on the TIP problem with Coulomb repulsion in two dimensions. 
We summarize our findings as follows: 
(1) we numerically compute TIP wavefunctions and give a direct evidence that 
the Coulomb interaction leads to the delocalization of  excited states;    
(2) we show that with the increase of the excitation energy
the level spacing statistics $P(s)$ exhibits
a transition from the Poisson distribution to the
Wigner-Dyson distribution and that at the critical point $P(s)$
is similar to the critical statistics found in the
3D Anderson model with broken time reversal symmetry \cite{zhar}. 

The paper is organized as follows. The model is introduced in Section 
\ref{model}. In Section \ref{estimates} we review the analytical 
arguments developed \cite{dima2d} for the TIP problem in two dimensions 
with Coulomb repulsion and we discuss the influence of a magnetic 
field on this theory.      
In Section \ref{deloc} and \ref{statistics} we discuss our numerical 
data for this problem when the time reversal symmetry is broken 
by a magnetic field. A number of   typical examples
of interaction induced pair delocalization
is shown in Section \ref{deloc}. The transition in the level spacing statistics
from the Poisson distribution to the Wigner-Dyson distribution
is analysed in Section \ref{statistics}. There we present the comparison of our 
results with the
data for the 3D Anderson transition with time reversal symmetry 
broken by a magnetic field. In Section \ref{conc} we 
present the summary of the results.   

\section{The model}
\label{model} 

We consider two particles with Coulomb repulsion in a two-dimensional 
disordered square lattice, in the presence of a constant magnetic field 
perpendicular to the plane. 
We restrict our investigations to the triplet case, which corresponds to 
the study of two spinless fermions. The singlet case, investigated in one 
and two dimensions for the on-site Hubbard interaction \cite{cuevas}, 
should give similar results. 
The Hamiltonian of the model reads: 
\begin{equation} 
\label{hamiltonian} 
H=-\sum_{<{\bf r},{\bf r'}>} V_{{\bf r},{\bf r'}} c_{\bf r}^\dagger 
c_{\bf r'} +\sum_{\bf r} E_{\bf r} n_{\bf r} + H_{int}.  
\end{equation}  
The vectors ${\bf r}=(x,y)$ denote the $L\times L$ sites of a square 
lattice with periodic boundary conditions applied in both directions,    
$c_{\bf r}^\dagger$ ($c_{\bf r}$) creates (destroys) an electron  
in the site ${\bf r}$. 
The occupation number at the site ${\bf r}$ is 
$n_{\bf r}=c_{\bf r}^\dagger c_{\bf r}$. 
The uncorrelated random energies $E_{\bf r}$ 
are distributed with constant probability within the interval 
$[-W/2,W/2]$, where $W$ denotes the magnitude of the disorder. 
The nearest neighbors hopping terms on the square lattice include the 
magnetic field being
$V_{{\bf r},{\bf r'}}=V\exp(\pm i 2\pi \alpha y)$ for
${\bf r}-{\bf r'}=(\pm 1,0)$,  while for  ${\bf r}-{\bf r'}=(0,\pm 1)$
they are $V_{{\bf r},{\bf r'}}=V$. This choice corresponds to
the Landau gauge for the vector potential, 
${\bf A}=(-Bya,0,0)$, with the magnetic field $B$ perpendicular to 
the plane. The number of flux quanta per unit cell of the lattice 
is  $\alpha=eBa^2/h=Ba^2/\phi_0$ and in the following the 
lattice spacing constant $a$ is taken to be unity.         
The magnetic field is chosen to be commensurate with the lattice, 
i.e. $\alpha=k/L$, with $k$ integer. 
The last term in (\ref{hamiltonian}) gives the interaction: 
\begin{equation}
H_{int}=\frac{U}{2} \sum_{{\bf r}\neq {\bf r'}} 
\frac{n_{\bf r} n_{\bf r'}}{|{\bf r}-{\bf r'}|},  
\end{equation} 
where $U$ is the strength of the Coulomb repulsion and $|{\bf r}-{\bf r'}|$ 
is the interparticle nearest distance computed on a 2D torus.  

\section{Analytical Estimates} 
\label{estimates} 

We consider the case with the average distance between electrons 
$R=|{\bf r}_1-{\bf r}_2|$ much larger than their one-particle 
localization length: $R\gg l_1$.  
In the localized regime the one-body Anderson localized orbitals 
can be represented in the lattice basis as: 
\begin{equation} 
\label{ergodic} 
\phi_\alpha ({\bf r}) \approx \frac{1}{l_1} \exp 
\left(-\frac{|{\bf r}-\overline{\bf r}_\alpha|}{l_1}+
i\theta_\alpha({\bf r})\right),  
\end{equation} 
where $\overline{\bf r}_\alpha$ marks the center of the localized
$\alpha$-th single particle eigenstate, and 
$\theta_\alpha({\bf r})$ is a random phase. 
Interaction matrix elements between noninteracting two-particle 
eigenstates $|\alpha\beta\rangle$ and $|\gamma\delta\rangle$ are given by 
\begin{equation} 
\langle\gamma\delta|H_{int}|\alpha\beta\rangle= 
Q_{\alpha\beta}^{\gamma\delta}-     
Q_{\alpha\beta}^{\delta\gamma},  
\end{equation}      
with 
\begin{equation} 
Q_{\alpha\beta}^{\gamma\delta}=
U\sum_{{\bf r}\neq{\bf r}'} 
\frac{\phi_\alpha({\bf r})\phi_\beta({\bf r}')
\phi^\star_\gamma({\bf r})\phi^\star_\delta({\bf r}')}
{|{\bf r}-{\bf r}'|}. 
\end{equation}  

Due to one-particle exponential localization, Coulomb repulsion 
can induce electron jumps only inside the localization domain 
of size $l_1$. Therefore, when $R\gg l_1$, it is possible to 
expand the interaction 
for electron displacements $\Delta{\bf r}_1,\Delta{\bf r}_2$ of 
typical length $l_1$  
near their initial positions ${\bf r}_1,{\bf r}_2$. The terms up to 
the first order in the expansion of the Coulomb potential give only 
mean-field corrections to the one-particle potential. 
The first term beyond mean field has a dipole-dipole form, and is 
of the order of:
\begin{equation} 
U_{dd}\sim-\frac{U}{R^3} \Delta {\bf r}_1 \cdot \Delta {\bf r}_2 
\sim \frac{Ul_1^2}{R^3}.  
\end{equation} 
This gives dipole-dipole matrix elements between noninteracting 
eigenstates:   
\begin{equation} 
\label{dipole} 
(Q_{\alpha\beta}^{\gamma\delta})_{dd} \sim
-\frac{U}{R^3}\sum_{{\bf r}_1,{\bf r}_2} \Delta {\bf r}_1 
\cdot \Delta {\bf r}_2
\phi_\alpha({\bf r}_1)\phi_\beta({\bf r}_2)
\phi^\star_\gamma({\bf r}_1)\phi^\star_\delta({\bf r}_2). 
\end{equation}  
The sum in (\ref{dipole}) runs over $l_1^2$ sites for each electron, 
so that in total the sum contains of the order of $l_1^4$
terms with random signs. Each term is of the order of $l_1^2 \phi^4 \sim
l_1^{-2}$. As a result,
the typical dipole-dipole transition matrix element in the ergodic 
approximation and with eigenstates given by (\ref{ergodic}) 
is of the order of 
\begin{equation} 
Q_{dd}^{\rm typ}\approx \frac{U}{R^3}. 
\end{equation}  

On the basis of this result we can estimate the typical interaction induced
transition rate  
$\Gamma_e$ between  noninteracting 
two-particle eigenstates by means of the Fermi golden rule: 
\begin{equation} 
\Gamma_e\sim (Q_{dd}^{\rm typ})^2\rho_2\sim\frac{U^2 l_1^4}{R^6V} \; .
\end{equation} 
Here we took the density of states coupled by interaction,
in the middle of the energy band of width $B\sim V$, 
being $\rho_2\sim l_1^4/V$. Indeed,  due to localization, 
one-electron jumps on 
a distance larger than $l_1$ give exponentially small matrix elements
and these transitions can be excluded from consideration.    
The mixing of two-electron states takes place when  
\begin{equation} 
\kappa_e=\Gamma_e\rho_2 \sim \left(\frac{U l_1^4}{V R^3}\right)^2 > 1, 
\end{equation} 
that corresponds to $R<l_1(Ul_1/V)^{1/3}$. For $U\sim V$ one gets 
$R<l_1^{4/3}$, and the condition $R\gg l_1$ is still satisfied 
when $l_1\gg 1$ \cite{tip2}.  
Therefore the physical picture is qualitatively different from 
the case of short-range screened interaction, where mixing is 
possible only for states at a distance $R<l_1$.  
For $\kappa_e>1$ the pair jumps on a typical length $l_1$ and 
its diffusion  rate  is
\begin{equation} 
D_e\sim l_1^2 \Gamma_e \sim \frac{V \kappa_e}{l_1^2}.   
\end{equation}  

The transition from localization to 
pair diffusion takes place in a way qualitatively similar 
to the Anderson model in 3D. Indeed, the pair center of mass 
can move in the 2D plane and in addition the electrons diffusively 
rotate around it in a ring of radius $R$ and width $l_1$, keeping their 
Coulomb energy $E_{ee}\sim U/R$ constant.    
The number of effective sites in the third direction,   
$M_{ef}\approx \pi R/l_1$, is given by the number of circles of 
size $l_1$ in the ring. 
Therefore, following standard 
results for the quasi-2D Anderson model \cite{lee}, the    
pair localization length $l_c$ is given by 
\begin{equation} 
\frac{l_c}{l_1}\sim\exp(M_{ef}g_2)\sim\exp
\left(\frac{\pi R\kappa_e}{l_1}\right),
\end{equation}  
where $g_2\sim\kappa_e$ is the two-particle conductance \cite{imry} 
and the above estimate is valid in the metallic phase for the 
corresponding 3D Anderson model ($\kappa_e>1$).
Since $R\sim l_1^{4/3}$ when $\kappa_e\sim 1$ (for $U\sim V$),
at the transition the TIP localization length jumps from 
$l_c\sim l_1$ to an exponentially large value  
\begin{equation} 
l_c \sim l_1\exp(\pi l_1^{1/3}).  
\end{equation}  
The TIP diffusion will be eventually localized due to the finite 
number of planes in the third direction. However, if disorder is 
not too strong ($l_1\gg 1$), Coulomb interaction gives raise to an 
exponentially sharp localization length enhancement, with a 
``critical'' behavior similar to the 3D Anderson model for finite 
system sizes $l_c\gg L \gg l_1$. 
We remark that, due to disorder, the ring is not rigid and
the electronic motion adapts to the fluctuations of the random 
potential so that the total energy remains constant even with 
some variation of the ring radius. This effect should increase 
the number of planes $M_{ef}$ of the effective quasi-2D Anderson 
model, giving a stronger delocalization effect. 

Since the excitation energy $\epsilon$ is related to the pair 
distance, $\epsilon\sim U/R$, the condition of pair delocalization 
($k_e>1$) implies 
\begin{equation}  
\epsilon>\epsilon_c\propto l_1^{-4/3};  
\end{equation} 
the scaling relation $\epsilon_c l_1^{4/3}={\rm const}$ is in agreement with 
the numerical results obtained  in Ref. \cite{dima2d}. 

\begin{figure} 
\centerline{\epsfxsize=8cm\epsffile{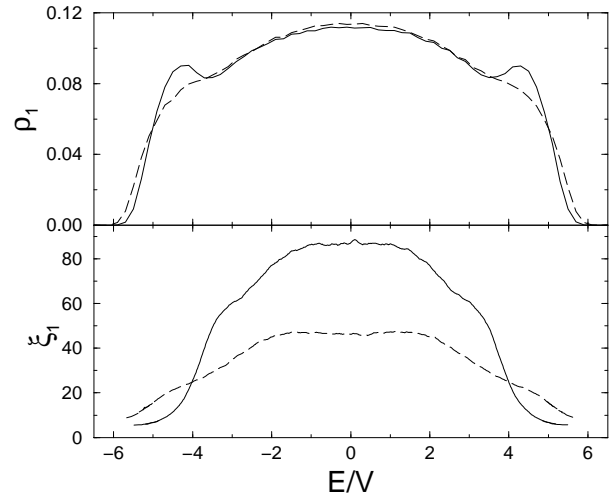}}
\caption{Single particle density of states $\rho_1$ 
(top) and the inverse participation ratio
$\xi_1\sim l_1^2$ (bottom) in the 2D Anderson model as a function 
of energy, for system size $L=24$, disorder strength $W=7V$, rescaled 
magnetic field $\alpha=1/6$ (full line) and $\alpha=0$ (dashed 
line). Data are averaged over $N_R=10^3$ disorder realizations. 
}    
\label{fig1} 
\end{figure} 

Finally we discuss the influence of a magnetic field on the 
theory presented in this Section. 
For a typical  magnetic field corresponding to $\alpha=1/6$ 
flux quanta per plaquette, the disorder strengths $W=7V$ or $W=10V$ is
strong enough to mix different Landau levels, as it is illustrated  in 
Fig.\ref{fig1} top: the single-particle density of states at 
$W=7V$ is only slightly changed with respect to the zero magnetic 
field case.     
In Fig.\ref{fig1} bottom we evaluate the energy 
dependence of the single particle  inverse participation ratio $\xi_1$ (IPR): 
\begin{equation} 
\xi_1=1/\sum_{\bf r} |\phi({\bf r})|^4, \quad \l_1\sim\sqrt{\xi_1} \; . 
\end{equation} 
This convenient characteristic $\xi_1$ determines how many sites contribute
to an eigenstate and is simply related to the localization length $l_1$.
The magnetic field gives an increase of $\xi_1$ in the 
middle of the energy band. Indeed, it is known from weak 
localization \cite{lee} that, in the presence of a magnetic 
field, coherent time-reversed paths are eliminated and therefore 
backscattering is suppressed. On the contrary, the magnetic 
field shrinks the band and at $W=7V$ there is a slight reduction of
the density of states near the band edges, which brings about 
a decrease in $l_1$ \cite{zhar3}.  For the parameters of Fig.\ref{fig1} 
the results found in \cite{dima2d} give the
critical delocalization  energy $\epsilon_c \approx 1.2 V$,
counted from the energy of the ground state 
(Fig.1c there), that corresponds to $E \approx -3.8 V$ in Fig.\ref{fig1}.
At this energy $\xi_1(\alpha=1/6)\approx \xi_1(\alpha=0)$ (see 
Fig.\ref{fig1} bottom)
that implies $l_1(\alpha=1/6)\approx l_1(\alpha=0)$ and hence 
we expect that the critical energy for delocalisation transition
in a presence of a magnetic field remains approximately the same:
$\epsilon_c(\alpha=1/6)\approx 
\epsilon_c(\alpha=0)$ \cite{energy}.  

\begin{figure} 
\vspace{-0.5cm} 
\centerline{\epsfxsize=4.2cm\epsffile{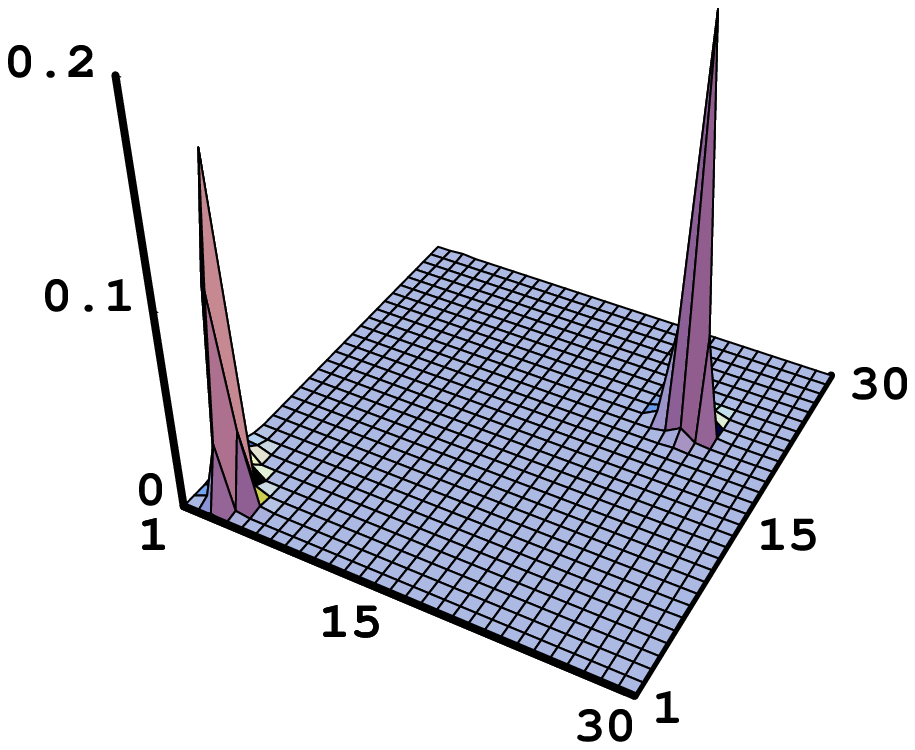}
\hfill\epsfxsize=4.2cm\epsffile{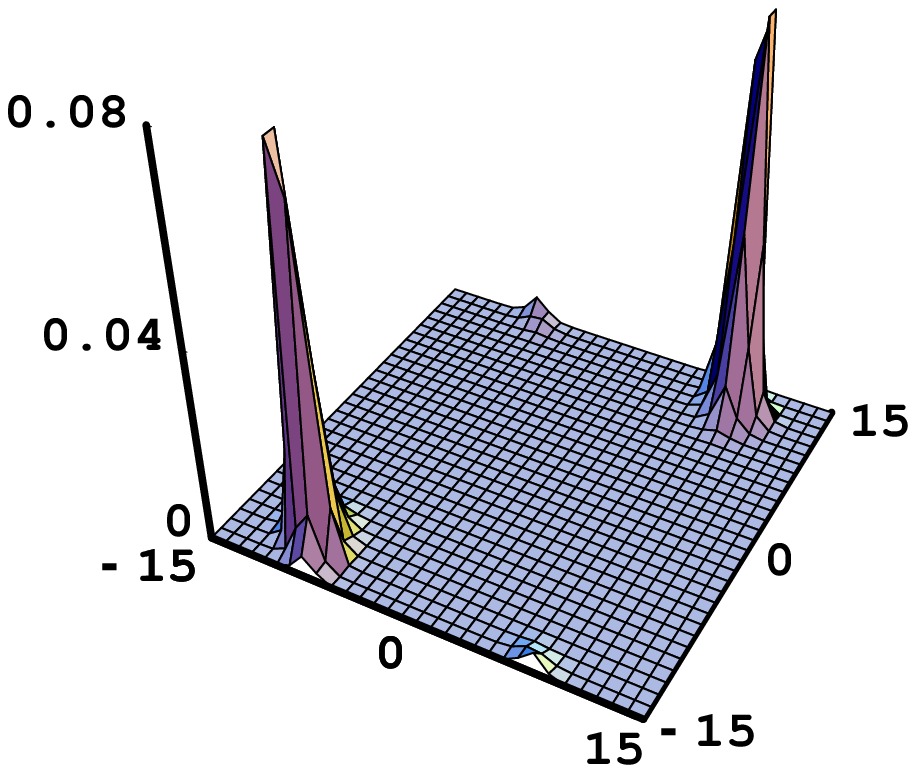}} 
\vspace{-1.cm} 
\centerline{\epsfxsize=4.2cm\epsffile{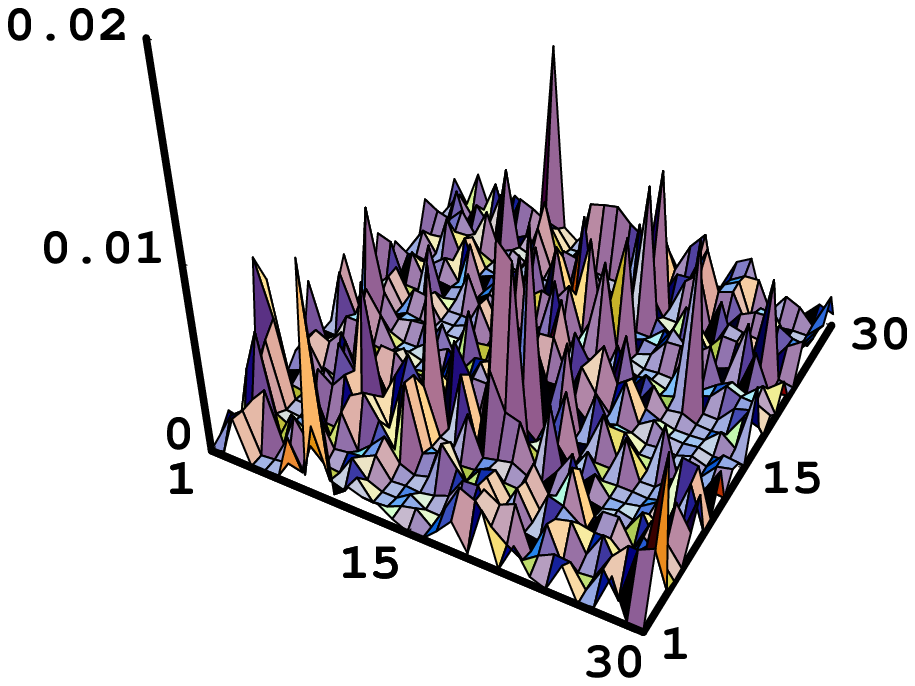}
\hfill\epsfxsize=4.2cm\epsffile{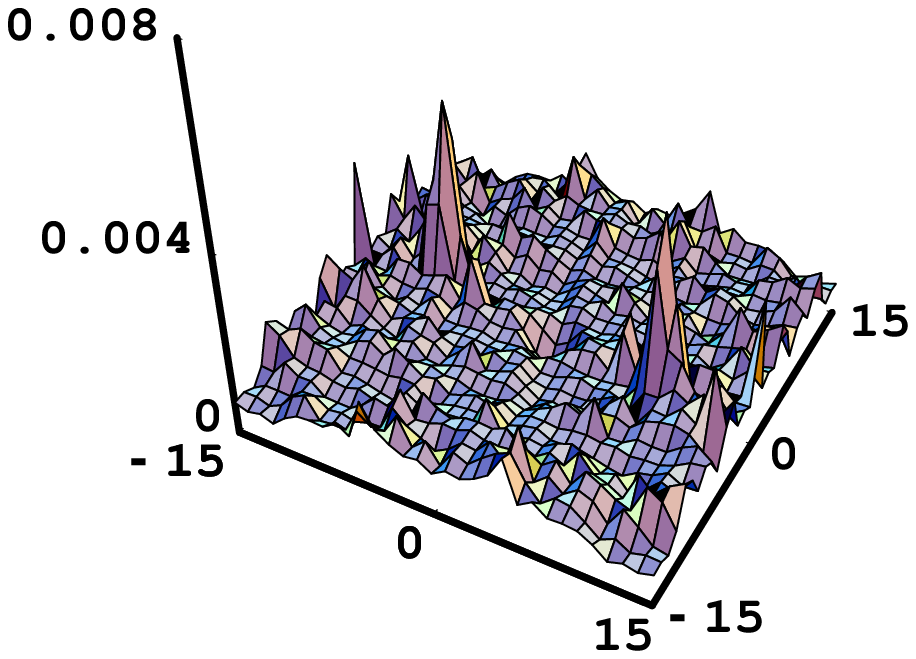}} 
\vspace{-1.cm} 
\centerline{\epsfxsize=4.2cm\epsffile{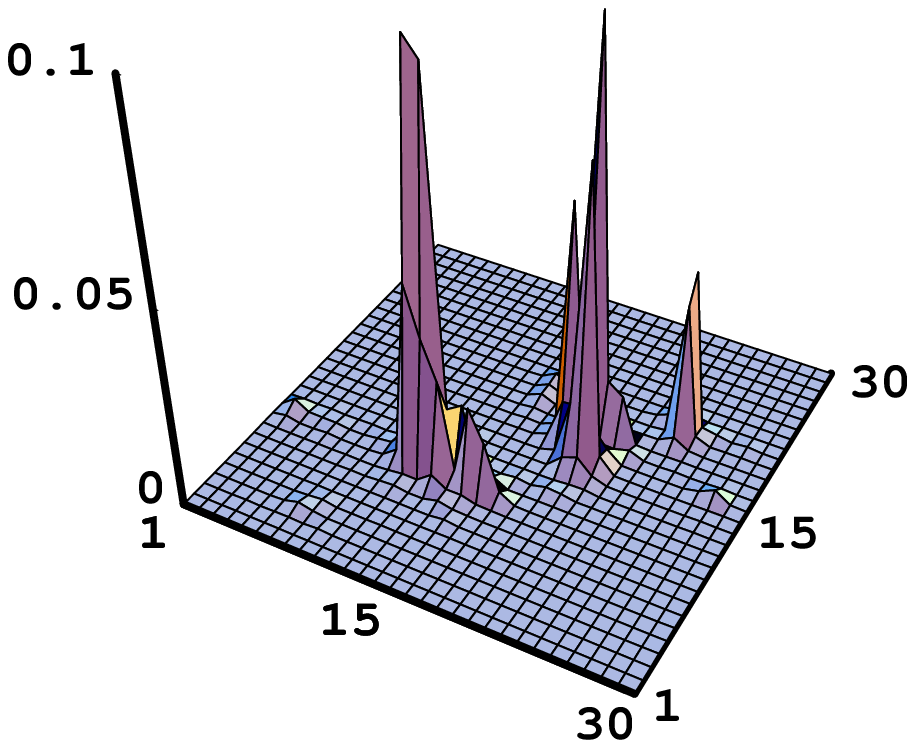}
\hfill\epsfxsize=4.2cm\epsffile{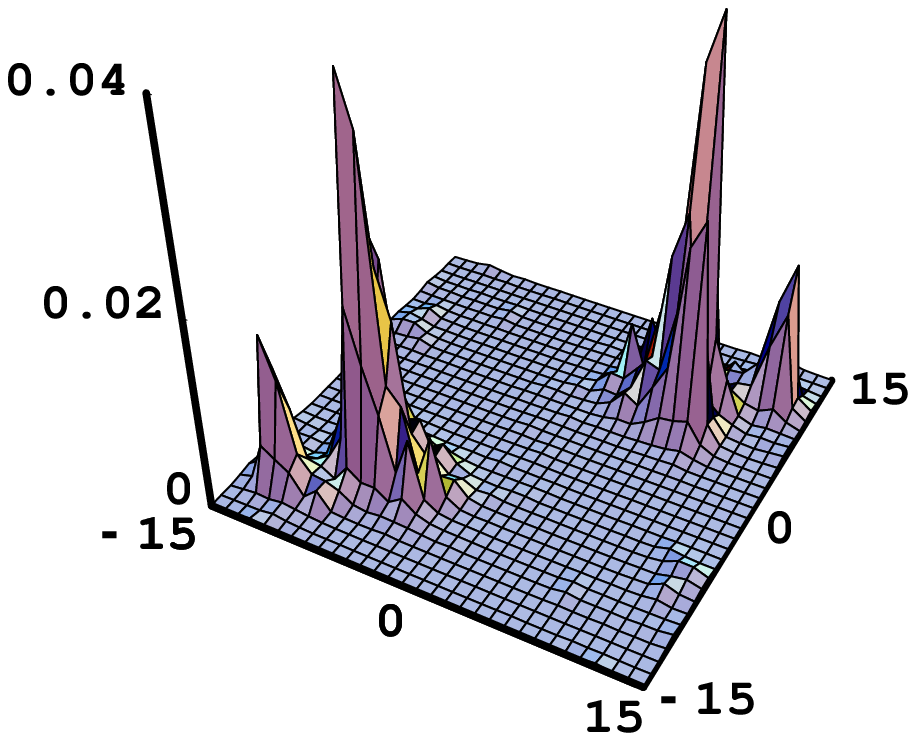}} 
\caption{
Probability distributions $f$ (left) and 
$f_d$ (right) from Eqs. (16) and (17)
for two interacting particles in a 
2D lattice of size $L=30$, with disorder 
strength $W=10V$ and rescaled magnetic field $\alpha=1/6$.         
Top: Coulomb repulsion $U=2V$, ground state, 
number of sites occupied by one particle given by IPR is $\xi_s=8$; 
middle: $U=2V$, rescaled one-electron excitation energy 
$\epsilon/B=0.71$ ($B=4 V$), $\xi_s=320$; 
bottom: same excitation energy but at $U=0$, 
$\xi_s=14.7$.       
}    
\label{fig2} 
\end{figure} 

\section{TIP delocalization}
\label{deloc} 

In order to study the eigenstate properties of 
our model with interaction, we diagonalize numerically the Hamiltonian 
(\ref{hamiltonian}). 
In this way we determine the two-particle probability distribution 
$F_k({\bf r}_1,{\bf r}_2)=|\Psi_k({\bf r}_1,{\bf r}_2)|^2$ 
for the $k$-th eigenfunction $\Psi_k({\bf r}_1,{\bf r}_2)=\langle {\bf r}_1 
{\bf r}_2 | \Psi_k\rangle$ written in the lattice basis.   
>From this  we extract the one-particle probability,  
\begin{equation} 
f_k({\bf r}_1)=\sum_{{\bf r}_2} F_k({\bf r}_1,{\bf r}_2), 
\end{equation}  
and the probability of interparticle distance,  
\begin{equation} 
f_{dk}({\bf R})=\sum_{{\bf r}_2} F_k({\bf R}+{\bf r}_2,{\bf r}_2),
\end{equation}  
with ${\bf R}={\bf r}_1-{\bf r}_2$.  

\begin{figure} 
\centerline{\epsfxsize=4.2cm\epsffile{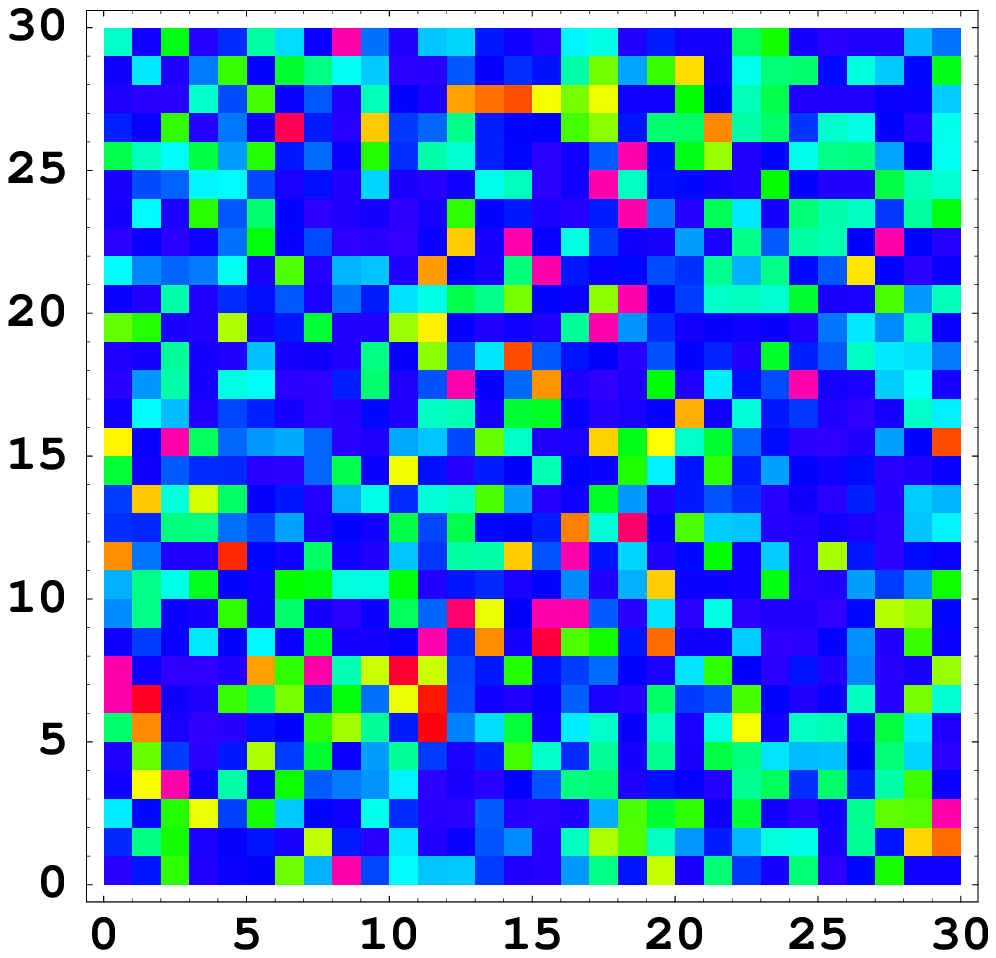}
\hfill\epsfxsize=4.2cm\epsffile{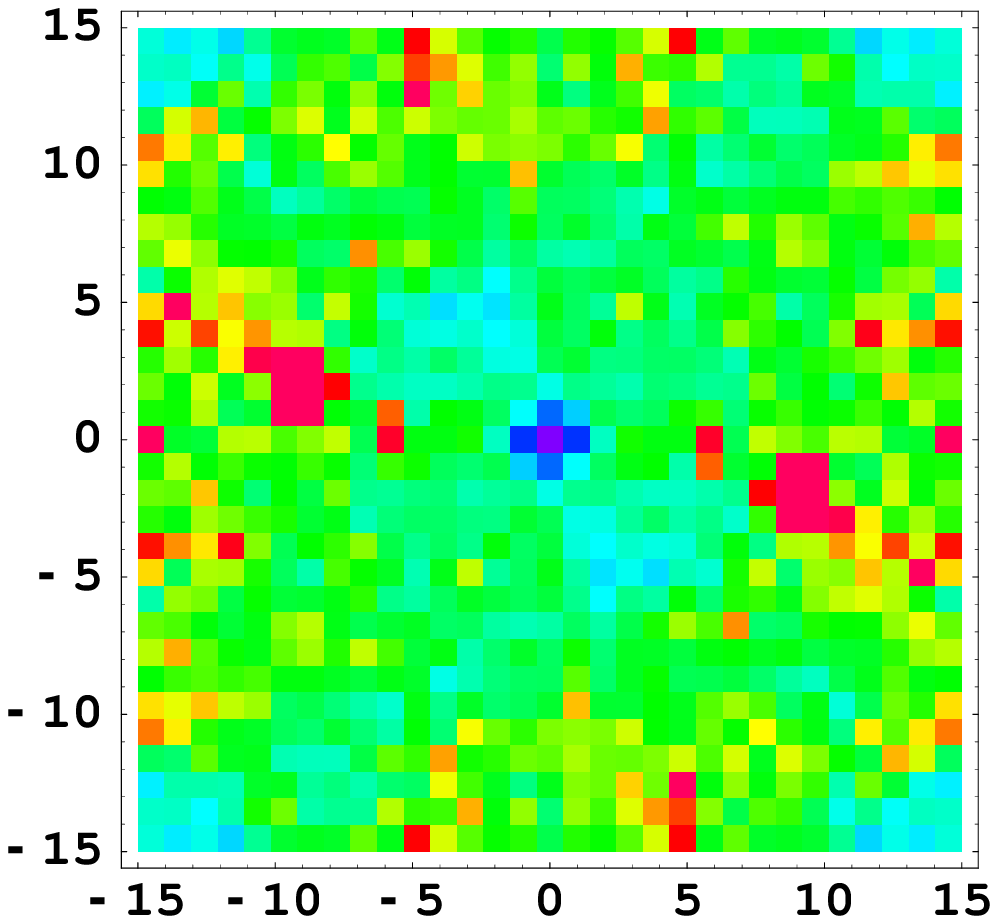}} 
\centerline{\epsfxsize=4.2cm\epsffile{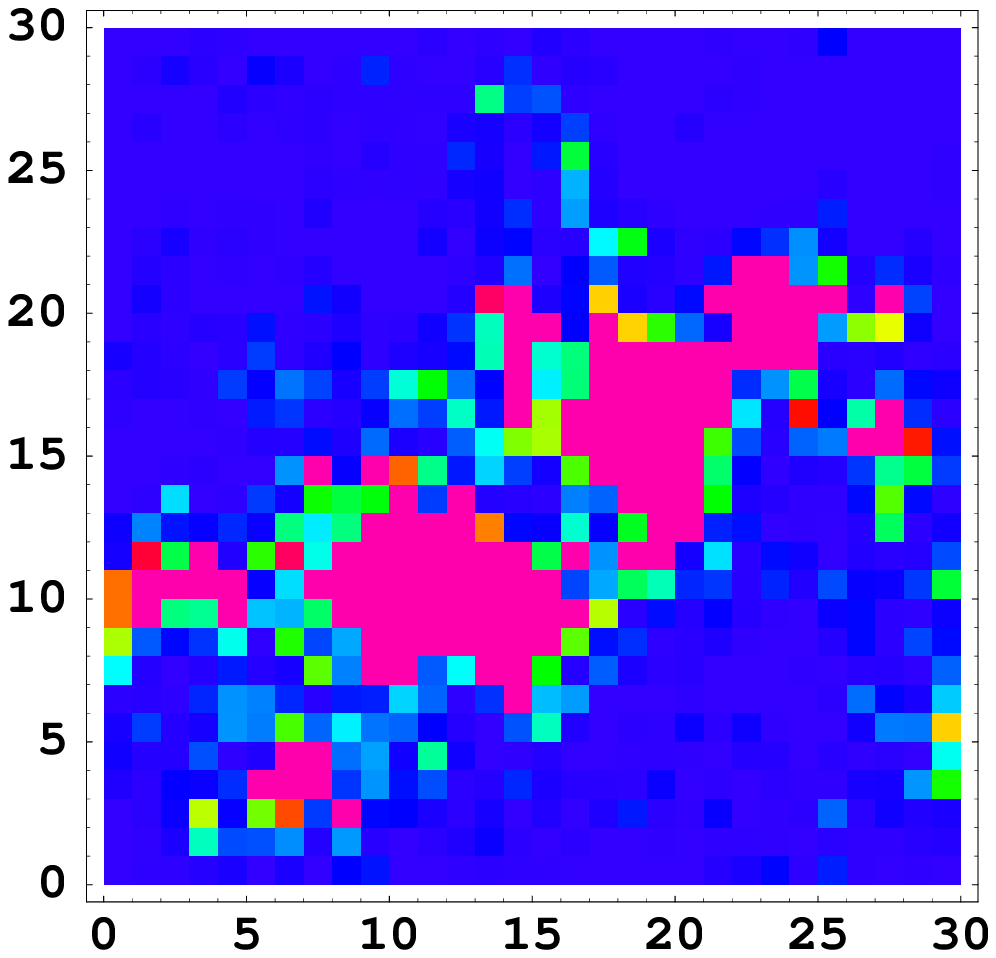}
\hfill\epsfxsize=4.2cm\epsffile{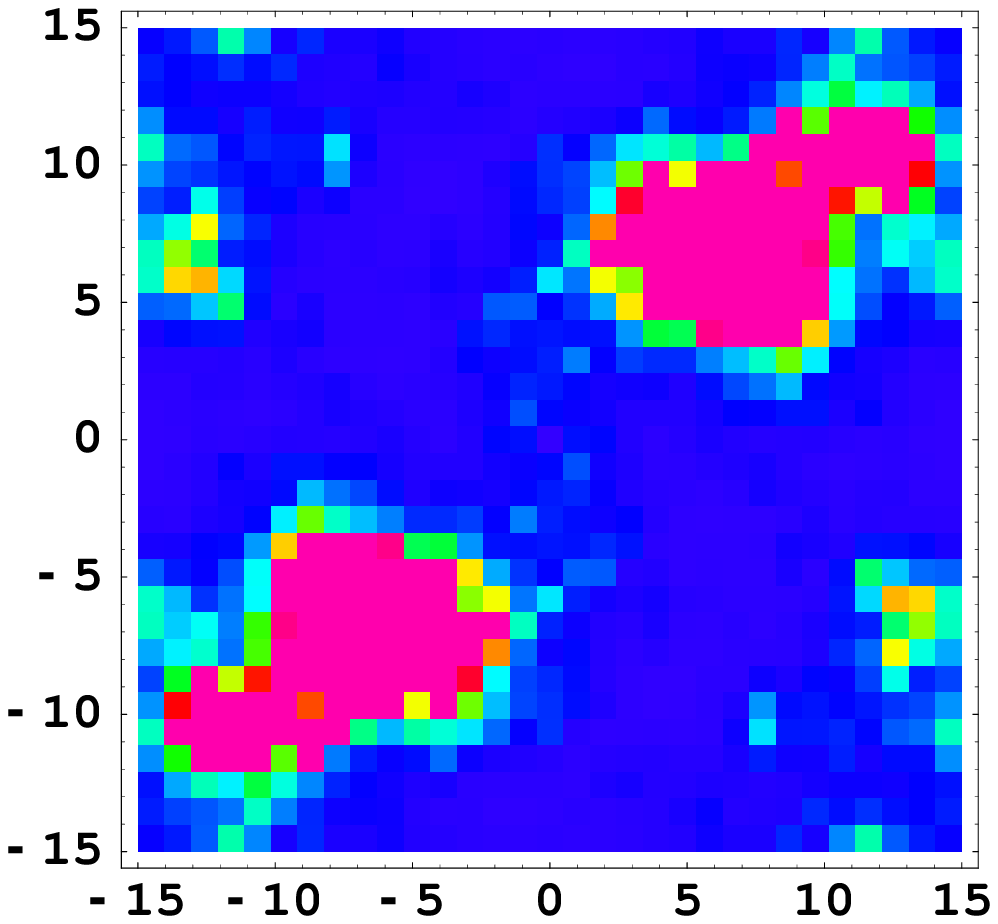}} 
\vspace{+0.5cm} 
\caption{Density plots for the case of Fig.\ref{fig2} middle
corresponding to two top plots ($U=2V$) and for the case
of Fig.\ref{fig2} bottom corresponding to two bottom plots ($U=0$).
Blue corresponds to the minimum of the probability distribution 
ad red to the maximum. 
}    
\label{fig3} 
\end{figure}

Typical examples of probability distributions are shown if 
Fig.\ref{fig2}, at $W=10V$ for a system size $L=30$. 
They clearly show that the two-particle ground 
state (Fig.\ref{fig2} top) remains localized in the presence of 
interaction, with the particles sitting far from each other in 
order to minimize Coulomb repulsion. Similar conclusions apply 
to low-energy eigenstates. On the contrary, for higher excitation 
energies ($\epsilon/B=0.71$ in Fig.\ref{fig2} middle, where 
$\epsilon=\delta E/2$, the excitation energy of 
the TIP eigenstate  $\delta E$ is counted from the ground state
and $B=4V$ is the band width at $W=0$) 
the probability distribution $f$ spreads over 
the whole lattice, while $f_d$ shows a hole at small $R$ and a depletion 
for large $R$. The first property is a simple consequence of Coulomb 
repulsion, while the second one is in agreement with the general 
discussion of the model (\ref{hamiltonian}). 
Following the analytical arguments \cite{dima2d} summarized in 
Section \ref{estimates}, we believe that this ring structure 
(see also Fig.\ref{fig3} top) 
would become more evident at larger system sizes, with maximum 
interparticle distance $R_{max}\approx L/2\gg l_1^{4/3}$ \cite{jose}.   
It is impossible to fully satisfy such a condition within the system 
sizes numerically tractable, if one considers that the condition 
$l_1\gg 1$ should be satisfied at the same time \cite{lanczos}.  
We stress that this pair delocalization takes place in a regime of 
strong localization for the one-particle wavefunctions. This 
is demonstrated in Fig.\ref{fig2} bottom and Fig.\ref{fig3} bottom, 
which show the 
probability distributions for the noninteracting problem 
($U=0$) at the same excitation energy. 
As a quantitative measure of the interaction induced charge 
delocalization one can take the inverse participation ratio 
$\xi_s$ for the one-particle probability $f$:  
\begin{equation} 
\xi_s=1/\sum_{{\bf r}_1} f^2 ({\bf r}_1). 
\end{equation} 
In this way $\xi_s$ gives the number of lattice sites occupied by one  
particle in an eigenstate. 
For the case of Fig.\ref{fig2}, Coulomb interaction does not 
significantly change $\xi_s\approx 8$ for the ground state, while 
for an excitation energy $\epsilon/B=0.71$ there is a huge delocalization 
effect from $\xi_s=14.7$ at $U=0$ to $\xi_s=320$ at $U=2V$. 

\begin{figure} 
\centerline{\epsfxsize=8cm\epsffile{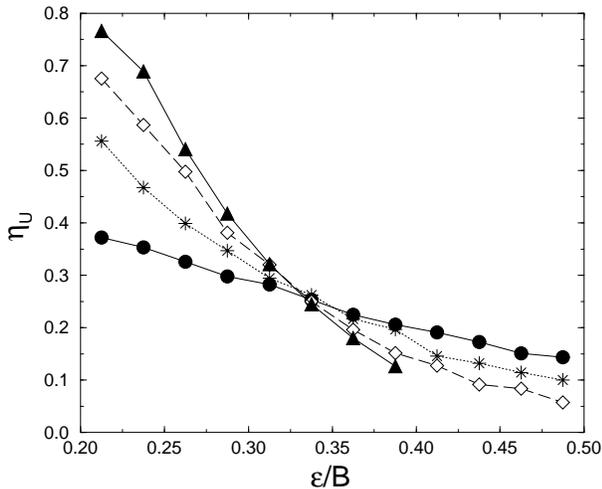}}
\caption{Dependence of $\eta_U$ on the rescaled one-electron energy 
$\epsilon/B$ for $W=7V$, $\alpha=1/6$, $U=2V$, system sizes 
$L=6$ (circles, number of disorder realizations $N_R=2\times 10^4$, 
number of spacings for each point of the graph $N_S>4\times 10^4$), 
$L=12$ (stars, $N_R=5\times 10^2$, $N_S>1.1\times 10^4$), 
$L=18$ (diamonds, $N_R=10^2$, $N_S>1.4\times 10^4$), 
and $L=24$ (triangles, $N_R=10$, $N_S>3\times 10^3$). 
}    
\label{fig4} 
\end{figure} 

\section{Spectral Statistics}
\label{statistics} 

The qualitative change of the structure of the eigenstates  
also leads to a change in the level spacing statistics. 
In the one particle problem spectral fluctuations proved to be 
a very useful tool to characterize the 3D Anderson transition 
\cite{shklovskii}. 
Localized wavefunctions yield uncorrelated spectra with 
Poisson statistics, characterized by a distribution $P(s)$ of 
the energy spacings between successive levels going to 
\begin{equation} 
P_P(s)=\exp(-s) 
\end{equation} 
when the system size $L\gg l_1$. Delocalized wavefunctions 
yield correlated spectra and Wigner-Dyson statistics 
with the Wigner surmise
\begin{equation} 
P_U(s)=\frac{32 s^2}{\pi^2}
\exp\left(-\frac{4 s^2}{\pi}\right) 
\end{equation} 
in the absence of time-reversal symmetry. 
The striking advantage of such an approach is that it deals only 
with the spectrum and does not involve heavy numerical calculations 
of conductivity or  eigenfunctions.   

\begin{figure} 
\centerline{\epsfxsize=8cm\epsffile{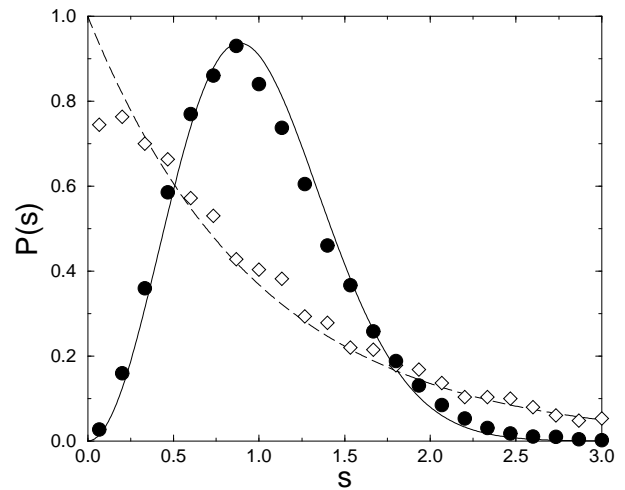}}
\caption{Transition from the Poisson
distribution (dashed curve)
to the Wigner surmise (full curve) in the level statistics 
$P(s)$ for the case of Fig.\ref{fig4} at $L=18$: $0<\epsilon/B<0.1$ 
(diamonds, $\eta_U=0.89$, $N_S=4.5\times 10^3$) 
and $0.47<\epsilon/B<0.5$ (circles, $\eta_U=0.06$, $N_S=3.5\times 10^4$). 
}    
\label{fig5} 
\end{figure} 

To analyze the evolution of the $P(s)$ distributions with 
respect to the excitation energy, it is convenient to use 
the parameter 
\begin{equation} 
\eta_U=\frac{\int_0^{s_U}[P(s)-P_U(s)]ds}
{\int_0^{s_U}[P_P(s)-P_U(s)]ds},
\end{equation}  
where $s_U=0.5076...$ is the first intersection point of $P_P(s)$ 
and $P_U(s)$. In this way $P_P(s)$ corresponds to $\eta_U=1$ 
and $P_U(s)$ to $\eta_U=0$. 
The dependence of $\eta_U$ on the one-electron excitation energy 
$\epsilon=\delta E/2$ 
is shown in Fig.\ref{fig4}. This figure 
shows that at fixed interaction $U=2V$, disorder $W=7V$, and 
rescaled magnetic field $\alpha=1/6$, curves at different system 
sizes $6\leq L \leq 24$ intersect at $\epsilon_c/B\approx 0.33$, 
with $\eta_{Uc}\approx 0.26$. 
While for $\epsilon<\epsilon_c$ the level spacing statistics 
approaches the Poisson limit ($\eta_U\to 1$) when the system 
size increases, for $\epsilon>\epsilon_c$ the tendency is 
towards the Wigner-Dyson distribution ($\eta_U\to 0$). 
We note that the critical excitation energies at $W=7V$ 
and $W=10V$ ($\epsilon_c\approx 0.67$, data not shown) are similar 
to the values found in \cite{dima2d}, in agreement with the 
the expectations of Section \ref{estimates}. 

\begin{figure} 
\centerline{\epsfxsize=8cm\epsffile{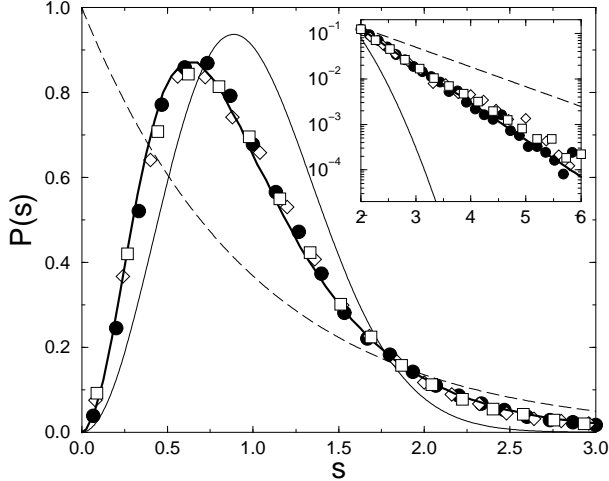}}
\caption{
Level statistics $P(s)$ at the critical point, for  
$U=2V$ and $\alpha=1/6$: $W=7V$, $0.32<\epsilon/B<0.35$, 
$L=6$ (circles, $N_S=8\times 10^4$) and $L=18$ 
(diamonds, $N_S=2\times 10^4$); $W=10V$, 
$0.65<\epsilon/B<0.70$, $L=12$ 
(squares, $N_R=2\times 10^4$, $N_S=10^5$). The thick line 
shows the critical $P(s)$ in the 3D Anderson model in the 
presence of a magnetic field (data taken from Fig.1 of Ref. [16]), 
with a large-$s$ exponential decay $P(s)\propto \exp(-\kappa s)$, 
$\kappa=1.87$, shown in the inset.   
}    
\label{fig6} 
\end{figure} 

The level spacing statistics $P(s)$ is shown in Fig.\ref{fig5} 
near the ground state and for excitation energies $\epsilon>
\epsilon_c$, for a system size $L=18$. The transition from the 
Poisson to the Wigner-Dyson statistics is evident. 

The $P(s)$ statistics near the critical point $\epsilon=\epsilon_c$ 
is shown in Fig.\ref{fig6}. The curves at different system sizes 
display a size-independent intermediate distribution, which exhibits 
level repulsion $P(s)\propto s^2$ at small $s$ and a Poisson-like 
tail $P(s)\propto \exp(-a s)$, with $a\approx 1.9$. 
The close agreement between these distributions and the critical 
statistics found in the 3D Anderson model with broken time 
reversal symmetry at the mobility edge, taken from Ref. \cite{zhar}, 
supports the analytical arguments given in Section \ref{estimates}. 
This 3D critical statistics also gives a good approximation for 
the 2D critical statistics at $W=10V$ (see Fig.\ref{fig6}), with 
$\eta_{Uc}\approx 0.29$. The small difference in the $\eta_{Uc}$ 
values at $W=7V,10V$ (already observed at $\alpha=0$, see 
Ref. \cite{dima2d}) could be attributed to the fact that the 
jump in the localization length at the ``transition'', 
$l_c/l_1\sim\exp(\pi l_1^{1/3})$, is not sufficiently sharp. 
This is due to the not very large values of $l_1$
accessible for numerical simulations.
Indeed, the investigation of cases with larger $l_1$ 
would require a significant increase of the system size, in order 
to satisfy the condition $L>R\approx l_1^{4/3}$. 
We also note that the finite statistics and the limited system 
sizes prevent us from precisely evaluating the critical excitation 
energy $\epsilon_c$ and the critical value $\eta_{Uc}$.  

\section{Conclusions}
We have shown that Coulomb interaction can delocalize electron pairs 
in a 2D disordered potential in the presence of a magnetic
field, above a critical excitation energy, in a way similar 
to the Anderson transition in 3D. 
The close relation  between these two transitions is reflected in 
the close similarity of level statistics at the critical point.  
The results obtained in this papers should be relevant for 
experiments at small electron density and/or large 
disorder fluctuations, when the distance between electrons is
larger than the size of their localization length 
in the absence of interaction.

\label{conc}

We thank Isa Zharekeshev for the possibility to use the data 
of Ref. \cite{zhar} ,   
and the IDRIS in Orsay and the CICT in Toulouse for access to 
their supercomputers.

\end{multicols} 


\begin{thebibliography}{99}
\bibitem{energy} A more detailed theory should take into 
account the variation of the transition rate $\Gamma_e$ and 
of the density of directly coupled states $\rho_2$ with 
the excitation energy $\epsilon$.
\end{thebibliography}
\end{document}